\documentclass[aps,showpacs,preprintnumbers,amsmath,amssymb, 
eqsecnum,
twocolumn, tightenlines,
]{revtex4}

\usepackage[dvips]{graphicx}
\usepackage{bm}
\begin{document}

\bibliographystyle{apsrev} 

\title {Amplitudes of\\
radiative corrections 
in fermion bags bound by Higgs boson exchange}

\author{M.Yu.Kuchiev} \email[Email:]{kmy@phys.unsw.edu.au}

%\author{V. V. Flambaum} \email[Email:]{flambaum@phys.unsw.edu.au}

%
\affiliation{School of Physics, University of New South Wales, Sydney
  2052, Australia}

%\author{E. Shuryak} \email[Email:]{shuryak@tonic.physics.sunysb.edu}
%
%\affiliation{Department of Physics, State University of New York,
% Stony Brook, NY 11794, USA}

    \date{\today}

    \begin{abstract} 
Properties of amplitudes that describe radiative corrections in a bag of heavy fermions bound by the Higgs boson exchange are studied. Classes of amplitudes, in which the large fermion mass is canceled out and hence produces no enhancement for the radiative corrections are found. For fermions with masses in the region $400\lesssim m \lesssim 1000$ Gev all relevant amplitudes are found to possess this property. Correspondingly the radiative corrections for this range of masses are small. For very heavy fermions, $m>1000$ Gev, the processes described by diagrams with closed fermion loops are also mass-independent.
    \end{abstract}


    \maketitle

\section{Introduction}
    \label{intro}

The mass of a fermion in the Standard Model arises from its interaction with the Higgs field. The heavier the fermion, the stronger the interaction. A scalar, Yukawa-type nature of the Higgs field makes this interaction attractive. When fermion is sufficiently heavy it can strongly modify the scalar field in its vicinity. As a result creation of a bag made from several fermions becomes possible. Implications related to this phenomenon have been discussed for a long time
 \cite{%
Vinciarelli:1972zp,%
PhysRevD.9.2291,%
Chodos:1974je,%
Creutz:1974bw,%
%Creutz:1974nc,%
Bardeen:1974wr,%
Giles:1975gy,%
Huang:1975ih,%
PhysRevD.15.1694,%
PhysRevD.25.1951,%
PhysRevLett.53.2203,%
PhysRevD.32.1816,%
Khlebnikov:1986ky,%
Anderson:1990kb,%
MacKenzie:1991xg,%
Macpherson:1993rf}. 
Even for a single heavy fermion one can ask a question of whether it can exist in a `bag-state' in which the surrounding Higgs is diminished \cite{Johnson:1986xz}. One can approach the problem from the opposite perspective by considering a large number of participating fermions, which comprise the fermion bag.
%R. Johnson and J. Schechter, Phys. Rev. D36 (1987) 1484.
In the classical approximations both these scenarios seem to be allowed. However, it was shown in Refs. \cite{Dimopoulos:1990at,Bagger:1991pg,Farhi:1998vx,Farhi:2003iu} that quantum one-loop corrections may become repulsive and so large that they destabilize bags at both ends; the existence of the one-fermion bag was put in doubt in \cite{Dimopoulos:1990at,Farhi:2003iu}, while \cite{Bagger:1991pg} argued that the corrections strongly deflate the bag in the limit of large number of fermions.
%E. Farhi, N. Graham, R. L. Jae, V. Khemani and H. Weigel, Nucl. Phys. B 665, 623 (2003)
%[arXiv:hep-th/0303159].
%S. Dimopoulos, B. W. Lynn, S. B. Selipsky and N. Tetradis, Phys. Lett. B 253, 237 (1991)
%J. A. Bagger and S. G. Naculich, Phys. Rev. Lett. 67, 2252 (1991).

The authors of Refs. \cite{Froggatt:etal} suggested looking at the magic number, $12 = 6$ tops $+~6$ antitops occupying the lowest $S_{1/2}$ shell with 3 colors. Using model-type approach they estimated that such a system should form a tightly-bound state. 
More accurate calculations based on the nonrelativistic mean field  approach of \cite{Kuchiev:2008fd}, which were also supported by \cite{Richard:2008uq}, showed though that this 12 top-antitop system is unbound. For a  massless Higgs there exists a bound state of 12 tops, but it is bound weakly and fades out for realistic Higgs masses. This conclusion was also supported by the relativistic approach of Ref. \cite{Kuchiev:2010ux}, which showed that 12 heavy fermions can become bound only if the fermion mass exceeds the critical value, which equals $\sim 320-410$ Gev, where the range is due to the uncertainty of the Higgs mass $m_\text{H}\sim 100 - 200$ Gev. Thus, the hope to observe a bag constructed from a relatively small number, say 12 or less of heavy fermions, rests on the expectation that there exist heavy fermions of the next forth generation.  Another option for hunting for the bags bound by the Higgs field relies on the idea that a large number of different particles, W and Z bosons, tops, possibly some scalars, can comprise the bag \cite{Crichigno:2009kk,Crichigno:2010ky}, but we would not pursue this topic here.

It was shown in Ref. \cite{Kuchiev:2008gt} that fundamental properties of fermion bags constructed from a small number of heavy fermions are independent of the fermion mass. 
In particular, the mass and size of a bag are defined mostly by the Higgs VEV $246$ Gev. As a result the mass of the bag can be smaller than the mass of a single heavy fermion. This conclusion was supported by the mean field approximation. However, for heavy fermions one can anticipate that quantum corrections could be substantial. Basing estimates of these corrections on the results of \cite{Farhi:1998vx,Dimopoulos:1990at,Bagger:1991pg,Crichigno:2009kk} 
one is inclined to believe that these corrections lead to strong effective repulsion, which destabilizes such bags. 

One should keep in mind though that the mentioned works considered the systems different from the bag of several heavy fermions discussed in \cite{Kuchiev:2010ux,Kuchiev:2008gt}, dealing either with a single fermion problem \cite{Dimopoulos:1990at,Farhi:2003iu}  or with a bag of a large number of fermions \cite{Bagger:1991pg} or scalars \cite{Crichigno:2009kk}.  Meanwhile, there is no clear indication that the role played by radiative corrections is universal. Hence, in order to make an assertive conclusion about the size of radiative corrections in the bags discussed in \cite{Kuchiev:2010ux,Kuchiev:2008gt} one should consider these corrections precisely for that particular system of several heavy fermions. This paper makes a step in this direction  providing the necessary groundwork.

\section{Formulation of problem}
\label{Formulation}

Let us separate the scale of masses of heavy fermions, which comprise the bag into two regions. Firstly, consider fermions, which are `moderately heavy', having masses in the region of $400\lesssim m \lesssim 1000$ Gev. For this interval of masses a bag can certainly be produced \cite{Kuchiev:2010ux}, and at the same time one can rely on the fact that the Higgs field inside the fermion bag does not deviate significantly from its vacuum expectation value, and that the size of the bag is much larger than the fermion Compton radius. 

These two features simplify the problem. The considered fermion masses are sufficiently large to single out the Higgs-fermion interaction as the most important part of the problem. In the first approximation one can neglect therefore the gluons, $Z$ and $W$ bosons, {\em etc}.  Since the Higgs field inside the bag deviates slightly from its VEV  the propagators can be approximated by their vacuum expressions. The same argument allows one to restrict consideration to the Feynman diagrams with a minimal number of external Higgs legs, since each such leg brings into the amplitude a small factor proportional to the small deviation of the Higgs field from its VEV. The momentum transferred along these legs should be presumed small compared to the fermion mass, because the size of the bag is large. Similarly, since the fermion binding   is shallow the fermion external legs can be taken in the vicinity of the mass shell $p^2=m^2$.

\begin{figure}[htb]
\centering
\includegraphics[height=6.5 cm,keepaspectratio = true, 
%angle = -90
]{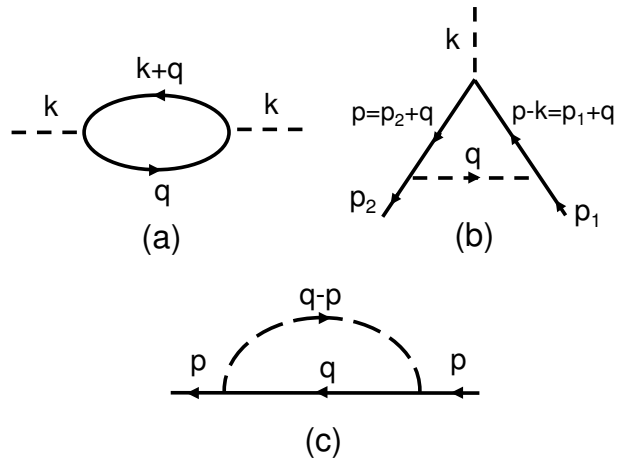}
\caption{
 \label{odin} Simplest one-loop corrections for fermions interacting with the Higgs field as specified in the Lagrangian (\ref{L0int})-(\ref{LY}). Solid and dashed lines describe fermions and Higgs bosons with propagators from (\ref{DG}) and vertexes from  (\ref{g}).}\end{figure}

These features point to a handful of amplitudes shown by diagrams (a), (b), and (c) in Fig. \ref{odin}, required for estimation of radiative corrections in the bag of several heavy fermions.  One of the purposes of this work is to present these amplitudes in the form convenient for future calculations.

The second, much more challenging region of interest, present `very heavy' fermions with masses above 1 Tev. Here one can approach the problem of radiative corrections using an interesting property of the solution found in \cite{Kuchiev:2008fd}. It was shown there that with an increase of the fermion mass the physical parameters of the fermion bag turn independent of this large mass. All fundamental properties of the bag, its size, its total energy {\it etc} are defined by the scale of the Higgs VEV, which equals  246 Gev, and which is well below the fermion mass. Consequently, in order to estimate the magnitude of 
the radiative corrections, the external momenta of all relevant diagrams 
can be taken as being small compared with the fermion mass, $|p_\mu|\ll m$. In the first approximation one can take zero momenta, $p^\mu=0$, for the external fermion legs. One of the aims of this work is to evaluate the necessary amplitudes in this limit. The difference with the case of the `moderately heavy' fermions is that these amplitudes are well outside the mass shell. We will restrict our discussion to the simplest one-loop approximation remembering though that at very large fermion masses this approximation is certainly not sufficient. The idea is that working out the one-loop corrections one can describe in simplest terms particular qualitative properties of the problem, which go beyond the limits of this approximation and would be useful in the following studies.

\noindent

Obviously, the perturbation theory based on the diagrams shown in Fig. \ref{odin} has been considered for different applications previously, see e.g. \cite{Bagger:1985ig,Babu:1985ut}. However, in order to make these diagrams useful for the problem at hand it is necessary to evaluate and summarize their properties paying particular attention to the kinematic regions mentioned above.

\section{Scalar-fermion interaction}
Consider fermions which interact with the Higgs field $\Phi$. Take  the conventional unitary gauge where $\Phi$ is represented by the real field $ \xi$
\begin{equation}
\Phi\,=\,\frac{v}{\sqrt{2}}\left( \begin{array}{c} 0 \\ \xi \end{array}\right)~.	
\label{Phi-xi}	
\end{equation}
Here $v=246$ Gev is the Higgs VEV. The part of the Lagrangian of the Standard Model, which describes interacting Higgs $\xi$ and fermion $\psi$ fields  reads 
($\hbar=c=1$)
\begin{align}
\!{\cal L}=
\frac{v^2}{2}\Big(\partial^\mu {\xi} \partial_\mu {\xi}\!-\!\frac{m_{\text H}^2}{4}(\xi^2\!-\!1)^2\Big)\!+
\bar \psi( i\gamma^\mu \partial_\mu \!-m \xi ) \psi\,.
\label{L}
\end{align}
We are interested in quantum fluctuations of the scalar field $\xi$ in the vicinity of its VEV, $\xi \approx \xi_\mathrm{VEV}=1$.
It is natural therefore to describe the Higgs field in terms of its deviation from the VEV $\varphi=v(\xi-1)$ rewriting the Lagrangian as follows
\begin{align}
&{\cal L}\,=\,{\cal L}_0+{\cal L}_{\text{NL}}+{\cal L}_{\text{Y}}~,
\label{L0int}
\\
&{\cal L}_0\,=\,
\frac{1}{2}\,(\,\partial^\mu {\varphi} \,\partial_\mu {\varphi}-m_{\text H}^2\,\varphi^2\,)+
\bar \psi\,( i\gamma^\mu \partial_\mu \!-m  ) \,\psi\,,
\label{L0}
\\
&{\cal L}_{\text{NL}}=-\frac{\lambda_3}{3!}\,\varphi^3 -\frac{\lambda_4}{4!}\,\varphi^4~,
\label{LNL}
\\
&{\cal L}_{\text{Y}}\,=\,-g\,\bar \psi \, \psi\,\varphi\,.
\label{LY}
\end{align}
Here the term ${\cal L}_0$ describes free propagation of non-interacting fermion and scalar fields with masses $m_{\text H}$   and $m$ and conventional propagators
\begin{equation}
D(p)\,=\,\frac{1}{p^2-m_{\text H}^2}~,\quad \quad G(p)\,=\,\frac{1}{\hat{p}-m}~.
\label{DG}
\end{equation}
The part ${\cal L}_{\text{NL}}$ of the Lagrangian 
describes the non-linear interaction of the Higgs field, while ${\cal L}_{\text{Y}}$ is responsible for the Yukawa-type interaction between the scalar and spinor fields. 
Eq. (\ref{L}) implies that the coupling constants, which appear in (\ref{LNL}) and (\ref{LY})  are 
\begin{equation}
\lambda_4\,=\,\frac{\lambda_3}{v}\,=\,3 \left(\frac{ m_{\text H} } {v}\right)^2~,
\quad\quad
g\,=\,\frac{ m} {v}~.
\label{g}
\end{equation}

\section{Vacuum polarization by fermions}
\label{Vacuum-polarization-fermions}

%\subsection{Vacuum polarization by fermions}

Consider the process shown in Fig. \ref{odin} (a), which describes the influence of the vacuum polarization by fermions on the Higgs field.  For the corresponding polarization operator $P(t)$ (self-energy correction for the Higgs), which describes correction to the propagator of the Higgs field, one can write
\begin{equation}
P(t)\,=\,-i \frac{g^2}{(2\pi)^4}\,\int\,\mathrm{Tr}\,\big(\,G(q)\,G(k+q)\,\big) \,d^4q~,
\label{P}
\end{equation}
where $t=k^2+i0$.  Conventional renormalization and straightforward calculations give
\begin{align}
P(t)=\frac{g^2}{8\pi^2}m^{2}\big(
F(w)\!-\!
F(\mu^2)\!-\!(w-\mu^2)F^{\,\prime}(\mu^2)\,\big).
\label{Pf}
\end{align}
Here useful notation 
\begin{align}
&w\,=\,t/m^2=k^2/m~, 
\label{wtk2}
\\
&\mu=m_{\text{H}}/m~,
\label{mu}
\end{align}
is introduced, and the function $F(w)$ is defined as follows
\begin{align}
&F(w)\,=\,2\,(w-4)\,\phi(w)~,
\label{F(w)}
\\
&\phi(w)\,=\,\Big(\frac{4-w}{w}\Big)^{1/2} \arctan \Big( \frac{w}{4-w}\Big)^{1/2}~.
\label{phi}
\end{align}
%(w-4)\int_0^1 \ln\,[\,1-w\,x\,(1-x)\,]\,dx+F_0(w)\\& \quad\quad =\,
%In the first line  of Eq.(\ref{F(w})  $F_0(w)$ refers to an arbitrary linear function of $w$, which has no impact on $P(t)$ being canceled out in Eq.(\ref{Pf}) anyway; to simplify notation  this function was specified to be $F_0(w)=2(w-4)$ in the second line.
To clarify notation it may be interesting to compare $P(t)$ with the well known one-loop expression for the polarization operator in spinor QED, see Appendix \ref{One-loop polarization operator in QED}.
The second and third terms in Eq.(\ref{Pf}) impose the renormalization conditions,
\begin{equation}
P(m_{\text{H}}^2)\,=\,P^{\,\prime}(m_{\text{H}}^2)=0~,
\label{ren}
\end{equation}
for the mass and wave function of the Higgs boson. 

It is instructive to consider the imaginary part of $P(t)$, which is present when $t>4m^2$.
In order to specify the necessary analytical continuation of
the elementary functions in Eqs.(\ref{F(w)}), (\ref{phi}) one notices that the rule $t\equiv
k^2+i0$ implies $(4-w)^{1/2}=-i(w-4)^{1/2}$ when $w>4$, and that
\begin{align}
\arctan \Big( \frac{w}{4-w}&\Big)^{1/2}\,=\,\frac{\pi}{2}-\arctan \Big( \frac{4-w}{w}\Big)^{1/2}
\nonumber
\\
&\,=\,\frac{\pi}{2}+i\,\mathrm{arctanh} \,\Big( \frac{w-4}{w}\Big)^{1/2}~.
\label{acr}
\end{align}
This transformation ensures that $\arctan[ {w}/{(4-w)}]^{1/2}$ is continuous at $w=4$.
The last identity presents the desired continuation for $w>4$.
From Eqs. ({\ref{P}), (\ref{F(w)}), (\ref{acr}) one finds that for $t>4m^2$
\begin{equation}
\mathrm{Im}\,P(t)\,=\,-\frac{g^2}{8\pi}\,\frac{(t-4m^2)^{3/2}}{\sqrt{t}}~.
\label{ImP}
\end{equation}
Let us  verify that this result complies with the unitarity condition. 
The latter can be written  as follows, see e.g. \cite{Berestetskii:1982},
\begin{equation}
 M_{if}-M_{fi}^*\,=\,\frac{1}{(4\pi)^2}\,
\frac{ | \bf p| }{ 2 \varepsilon }\,
\sum_{\sigma}\,\int M_{in}M_{fn}^*\,d\Omega_{\bf p}~.
\label{u1}
\end{equation}
Here the indexes $i,f$ and $n$ refer to the initial, final and intermediate states of a reaction. In the case considered the state $i=f$ describes the Higgs boson (virtual or real), while $n$ represent a pair of real fermions created by this boson. The momentum  of the fermion in the center of mass system (cms) is called ${\bf p}$, while its energy is $\varepsilon=\sqrt t/2$, the summation and integration in (\ref{u1}) run over allowed spin states and the angular distribution of the fermion pair. One takes into account that $M_{ii}=-P(t)$ and  $M_{in}=g\,\bar{u}(p{+})u(-p^{\,\prime})$, where $p$ and $p^{\,\prime}$ are momenta of the fermion and antifermion. The summation over their spin projections is fulfilled via the conventional identity $\sum_\sigma |M_{in}|^2=g^2\mathrm{Tr}\,
[ (\hat{p}+m)(\hat{p}^{\,\prime}-m)]=4g^2(\varepsilon^2+{\bf p}^2-m^2)$. The angular integration is reduced to a trivial factor of $4\pi$. Using also the kinematic conditions $\varepsilon=\sqrt{t}/2$ and $|{\bf p}|=(t-4m^2)^{1/2}/2$, one finds that the imaginary part of $P(t)$ derived from the unitarity condition (\ref{u1}) is identical to the one specified in (\ref{ImP}).

One can now verify the expression for $P(t)$ by re-deriving it from its imaginary part. One writes the dispersion relation for the second derivatives
\begin{equation}
P^{\,\prime\prime}(t)\,=\,\frac{1}{\pi}\int_{4m^2}^\infty \,
\frac{ \mathrm{Im}\,P^{\,\prime\prime}(\tau)}{\tau-t-i0}~d\tau~.
\label{P''}
\end{equation}
It is convergent and thus needs no additional subtractions.
Straightforward calculations show that thus defined second derivative
$P^{\,\prime\prime}(t)$ coincides with the one that follows from Eq. (\ref{Pf}).  Combining the dispersion relation (\ref{P''}) with the renormalization conditions (\ref{ren}) one fully recovers $P(t)$ in Eq.(\ref{Pf}) from the unitarity condition.

Consider behaviour of $P(t)$ in asymptotic regions. For large $|t|\gg m^2$ one finds from Eq.(\ref{Pf})
\begin{equation}
P(t)\,\approx\,
\frac{g^2}{8\pi^2}\,t\,\ln (\,-t/m^2-i0\,)~.
\label{Plarget}
\end{equation}
%where a conventional rule $t\equiv t+i0$ defines the value of the logarithmic function.
%\begin{equation}
%P(t)\,\approx\,
%\frac{g^2}{8\pi^2}\times\left\{
%\begin{array}{ll}
%t\,(\,\ln (\,t/m^2\,)-i\pi),& \quad t>0 
%\\
%t\,\ln (\,|\,t\,|/m^2\,),        &\quad t<0 
%\end{array}
%\right.
%\label{Plarget}
%\end{equation}
For small $|t|\ll m^2$ Eq. (\ref{Pf}) allows one to make an expansion
\begin{equation}
P(t)\,\approx\,\frac{g^2}{8\pi^2}\,m^2\,(p_0+p_1\,w+p_2\,w^2\,)~,
\label{small t}
\end{equation}
where, remember, $w=t/m^2$, while the expansion coefficients are
\begin{align}
&p_0\,=\,-8-F(\mu^2)+\mu^2\, F^{\,\prime}(\mu^2),
\\
&p_1\,=\,~\frac{8}{3}-F^{\,\prime}(\mu^2),\quad p_2\,=\,~\frac{1}{2}\,F^{\,\prime\prime}(\mu^2).
\end{align}
It was taken into account here that $F(0)=-8$, $F^{\,\prime}(0)=8/3$. 
For an interesting for applications case when the fermion mass $m$ is large, $m\gg m_\text{H}$, Eq.(\ref{small t}) can be simplified further
\begin{equation}
P(t)\,\approx\,
-\frac{g^2}{8\pi^2}\,\frac{(t-m_\text{H}^2)^2}{10\,m^2}\,=\,-\frac{1}{80\pi^2}\,\frac{(t-m_\text{H}^2)^2}{v^2}~.
\label{P2}
\end{equation}
The numerical coefficient $-1/10$ here follows from the identity $F^{\,\prime\prime}(0)=-1/5$.
From (\ref{P2}) we find, in particular, that at zero momentum $P(0)= -m_\text{H}^4/(80\pi^2v^2)$. 

Note a simple but important for applications fact. The fermion mass $m$ is canceled out in the final expression in Eq. (\ref{P2}). The cancellation takes place between the mass factor present in the coupling constant $g^2=m^2/v^2>1$ and a factor $m^2$ which arises in the denominator in the intermediate expression in (\ref{P2}) and is naturally explained by the renormalization conditions. If  $t$ and $m_\text{H}$ are both small, then the polarization operator describes those processes that are close to the mass shell. Here Eq.(\ref{ren}) makes it certain that only the second (and higher) derivative  of the polarization operator is essential. An expansion of $P(t)$ runs over the  ratios of $t/m^2$ and $m_\text{H}^2/m^2$. Thus the large fermion mass inevitably arises in the denominator. Similar cancellation of the fermion mass is found below for the vertex correction in Eq.(\ref{ft0}), and corrections produced by the fermion loop with several Higgs legs, see Eq.(\ref{3leg}).

Generally speaking, the cancellation of the fermion mass in $P(t)$ 
opens a way for discussion of the vacuum polarization produced by the loop of the Higgs field. However, from the physical point of view it is difficult to expect that this process gives  significant contribution. The calculations presented in Appendix 
\ref{Vacuum polarization by Higgs field} support this feeling.

\section{Yukawa formfactor}
\label{mass-shell}

Consider the process shown in Fig. \ref{odin} (b), which gives correction $\delta g$ to the vertex $g$ of the fermion-Higgs coupling. 
In this Section we restrict our discussion to the kinematic region, in which the incoming and outgoing fermions are on the mass shell, $p_1^2=p_2^2=m^2$. It was mentioned in Section \ref{Formulation} that this region is important for the fermion bags constructed from `moderately heavy' fermions.
In that case the vertex is a function of the only available variable, the transferred momentum $k$, $\delta g=\delta g(k)$ so that this correction can be considered as a formfactor related to the Yukawa charge. It is natural therefore to introduce notation 
\begin{equation}
g+\delta g(k)\,=\,g f_\text{Y}(k)~, 
\end{equation}
where $f_\text{Y}(k)$ is the mentioned formfactor. From diagram (b) one finds
\begin{equation}
f_\text{Y}(t)-1\,=\,\frac{i\,g^2}{(2\pi)^4}\int D(q)\,G(p_2+q)\,G(p_1+q)\,d^4q~,
\label{fY}
\end{equation}
where $t=k^2$.
Renormalization of the formfactor is achieved by ensuring that at small transfered momentum it exhibits a trivial value
\begin{equation}
f_\text{Y}(0)\,=\,1~.
\label{fY=1}
\end{equation}
Imposing this condition, one fulfills usual calculations of necessary integrals in Eq.(\ref{fY}). With this purpose the Feynman parametrization for three factors in the denominator of the integrand, which arise from the propagators (\ref{DG}), was employed.
After integration over $d^4q$, one is left with integrations over two auxiliary parameters introduced during the Feynman parametrization of the denominators. Integration over one of them is straightforward, while the remaining integral over the auxiliary parameter $0 \le x\le 1$ is convenient to keep in the final result, which reads
\begin{equation}
f_\text{Y}(t)-1\,=\,\frac{g^2}{8\pi^2}\int_0^1 \big(\, H(x,w)-H(x,0)\,\big)\,dx~,
\label{fYF}
\end{equation}
where according to (\ref{wtk2}) $w=t/m^2$ and the function $H(x,w)$ is defined as follows
%\begin{align}
%&H(x,w)\,=\,2\left(4\,(1-x)+\frac{m_\text{H}^2}{m^2}\right) \times
%\label{H}
%\\
%&\frac{1}{{s(x,w)\,\sqrt w}}\arctan \frac{x\,\sqrt{w}}{s(x,w)}+\frac{1}{2}\,\ln\big(1-w\,x(1-x)\big)~,
%\nonumber
%\end{align}
%
%
\begin{align}
&H(x,w)\,=\,2\big(4\,(1-x)+\mu^2\,\big) \times
\label{H}
\\
&\frac{1}{{Z(x,w)\,\sqrt w}}\arctan \frac{x\,\sqrt{w}}{Z(x,w)}+\frac{1}{2}\,\ln\big(1-w\,x(1-x)\big)~,
\nonumber
\\
&Z(x,w)\,=\,\big[\,(4-w)\,x^2+4\,\mu^2\,(1-x)\,\big]^{1/2}~.
\label{s(x)}
\end{align}
Here notation of (\ref{mu}) $\mu\,=\,{m_\text{H} }/{m}$ is used.
One anticipates that the integral in Eq.(\ref{fYF}) is calculable in elementary functions, but the outcome does not promise to be convenient. Fortunately, there is no urgent need for presenting it. For most applications the representation provided by (\ref{fYF}) is sufficient. Besides, below we introduce another useful expression for $f_\text{Y}(t)$, see Eq.(\ref{fYDR}).

For $t>4m^2$ there exists an imaginary part of the formfactor. The initial integral representation in (\ref{fY}) provides a convenient way of finding it. One rewrites (\ref{fY}) in a form
\begin{equation}
f_\text{Y}(t)-1\,=\,\int \frac{i\Phi(p)}{(p^2-m^2) \,((p-k)^2-m^2)}\,d^4p~,
\label{Phi}
\end{equation}
where $p=p_2+q$, $p-k=p_1+q$ and $\Phi(p)$ is defined as follows
\begin{equation}
\Phi(p)\,=\,\frac{g^2}{(2\pi)^4}\frac{(\hat q+2m)^2}{q^2-m_\text{H}^2}.
\label{Phi(p)}
\end{equation}
It was taken into account here that the external fermion legs are on the mass shell, which allows one to substitute $\hat p=\hat p_2+\hat q\rightarrow \hat q+m$, $\hat p - \hat k = \hat p_1+\hat q\rightarrow \hat q+m$,
and consequently
$(\hat p+m)(\hat p-\hat k+m)\rightarrow (\hat q+2m)^2$.

Eq.(\ref{Phi}) implies, compare section 117 of \cite{Berestetskii:1982}, 
\begin{equation}
2\, \mathrm{Im}\,f_\text{Y}(t)\,=\,-\frac{\pi^2}{2}\Big( \frac{t-4m^2}{t}\Big)^{1/2}\int \Phi(p)~d\Omega_{\boldsymbol{p}}~,
\label{2ImfY}
\end{equation}
where the integration is fulfilled in the center of mass coordinates for the two real fermions, which are present in the intermediate state, over their angular distribution. 
The integrand depends on the corresponding angles through the transferred momentum, which equals
$q=(0,{\bf p}-{\bf p}_2)$.
Substituting (\ref{Phi(p)}) in (\ref{2ImfY}) one finds after simple calculations
%\begin{align}
%&\mathrm{Im}\,f_\text{Y}(t)\,=\,\frac{g^2}{16\pi}\,\frac{1}{\sqrt{w\,(w-4)}}\times
%\label{ImfY}
%\\
%\nonumber
%&\left[ -w-4+\left(4+\frac{m_\text{H}^2}{m^2}\,\frac{w+4}{w-4}\right)\,
%\ln \left(1+\frac{m^2}{m_\text{H}^2}(w-4)\right)\right]~,
%\end{align}
\begin{align}
&\mathrm{Im}\,f_\text{Y}(t)\,=\,\frac{g^2}{16\pi}\,\frac{1}{\sqrt{w\,(w-4)}}\times
\label{ImfY}
\\
\nonumber
&\left[ -w-4+\left(4+\mu^2\,\frac{w+4}{w-4}\right)\,
\ln \left(1+\frac{w-4}{\mu^2}\right)\right]~,
\end{align}
where notation $w=t/m^2$ and $\mu=m_\text{H}/m$ is employed. 

The known imaginary part of the formfactor (\ref{ImfY}) provides one with an opportunity to present the formfactor itself via the following dispersion relation
\begin{equation}
f_\text{Y}(t)-1\,=\,\frac{t}{\pi}\,\int_{4m^2}^\infty\,\frac{\mathrm{Im}\,f_\text{Y}(\tau)}{\tau\,(\tau-t-i0)}\,d\tau~,
\label{fYDR}
\end{equation}
which is written here with one subtraction to accommodate the renormalization condition (\ref{fY=1}). 

To verify validity of these results the imaginary part of the formfactor was re-derived using the unitarity condition (\ref{u1}). 
The state $i$ in this condition is to be taken as a Higgs boson with the momentum $k$, the final state $f$ presents a fermion with momentum $p_2$ and antifermion with momentum $-p_1$, while the intermediate state $n$ is made of the fermion with momentum
$p=p_2+q$ and antifermion with momentum $k-p=-q-p_1$ . The pair of fermions in the intermediate state undergo the scattering due to exchange of the Higgs boson with momentum $q$. Correspondingly, $M_{if}=-g(f_\text{Y}(t)-1)$, $M_{in}=-g \bar u(p)u(p-k)$, and $M_{fn}=-g^2\,
[\bar u(p_2)u(p)][-\bar u(p-k)u(p_1)]/(q^2-m_\text{H}^2)$, where the sign minus in the second square brackets arises from the anifermion line. One verifies then that conventional summation over the spin projections in the unitarity condition brings it to the form identical to Eq.(\ref{2ImfY}). Subsequently the imaginary part derived from unitarity condition equals the one considered previously in Eq.(\ref{ImfY}). 

An additional verification was fulfilled using the two available representations for the formfactor provided by Eqs.(\ref{fYF}) and (\ref{fYDR}).  Numerical values of the formfactor, which were extracted from these two formulas, were verified  to agree for a wide range of parameters $t/m^2$ and $m_\text{H}^2/m^2$.

Consider asymptotic regions. For large $|t|\gg m^2,m_\text{H}^2$ one finds from Eqs.(\ref{ImfY}) and  (\ref{fYDR})
\begin{align}
&f_\text{Y}(t)-1\,\approx\,\frac{g^2}{16\pi^2}\,\big( \ln (-t/m^2-i0)+C
\,\big)~,
\\
&C=2+(4-3\mu^2)\ln \mu - 3\mu \sqrt{4-\mu^2}\,\arctan \frac{\sqrt{4-\mu^2}}{\mu},
\nonumber
\end{align}
%&C=2+\frac{4-3\xi}{2}\ln\xi-3\sqrt{\xi(4-\xi)}\,\arctan\Big(\frac{4-\xi}{\xi}\Big)^{1/2}.
where $\mu=m_\text{H}/m$. It is taken into account here that 
according to (\ref{ImfY}) $\mathrm{Im}\,f_\text{Y}(t)\rightarrow -g^2/(16\pi)$ when $t\rightarrow \infty$. This asymptotic value was singled out in the integral in (\ref{fYDR}), the subsequent  calculations were straightforward.

For small $ |t|\ll m^2 $ we find from Eq. (\ref{fYDR}) 
\begin{equation}
f_\text{Y}(t)-1\,\approx\,\frac{g^2}{8\pi^2}\,\frac{\gamma\,t}{m^2}\,=\,
\frac{1}{8\pi^2}\,\frac{\gamma\,t}{v^2}~.
\label{ft0}
\end{equation}
Note a simple and important fact. The dependence on the fermion mass $m$ is canceled out in the last identity here similarly to the way $m$ does not show itself in the polarization operator in Eq.(\ref{P2}). The reason is also similar as an $m$-dependence of the coupling constant $g=m/v$ is counter balanced by an $m$-dependence of the denominator, where $m^2$ appears due to expansion of the formfactor over the parameter $t/m^2$. (Speaking more precisely, only the  power-type dependence on $m$ is not present in (\ref{ft0}). Less pronounced $\propto \ln m $ behaviour is still present there, as $\gamma$ depends on $m$, see Eq.(\ref{gamma}) below. The point is that the cancellation of the large factor $m^2$ in Eq.(\ref{ft0}) reduces the dependence on $m$ significantly.)

According to Eq. (\ref{fYDR}) the coefficient $\gamma$ introduced in (\ref{ft0}) satisfies
\begin{equation}
\frac{g^2}{8\pi^2}\,\frac{\gamma}{m^2}\,=\,\frac{1}{\pi}\int_{4m^2}^\infty \frac{\mathrm{Im}\,f_\text{Y}(t)}{t^2}\,dt~.
\label{appear}
\end{equation}
Substituting $\mathrm{Im}\,f_\text{Y}(t)$ from (\ref{ImfY}) and calculating the resulting integral one finds that this coefficient equals
\begin{align}
\label{gamma}
\gamma\,=\,\frac{1}{12}\big[\,(18-7\mu^2)\,\phi(4-\mu^2)-(4-7\mu^2)\ln\mu-7\,\big]~,
%\frac{\mu}{\sqrt{4-\mu^2}}\, \arctan \frac{\sqrt{4-\mu^2}}{\mu}~\Big)~,
\end{align}
%
%\begin{align}
%\gamma\,=\,\frac{1}{12}&\Big[-7-\frac{4-7\xi}{2}\,\ln\xi\,+
%\label{gamma}
%\\
%&(18-7\xi)\,\Big(\frac{\xi}{4-\xi}\Big)^{1/2}\!\arctan \Big( \frac{4-\xi}{\xi}\Big)^{1/2} \,\Big]~,
%\nonumber
%\end{align}
where $\mu=m_\text{H}/m$ and $\phi(w)$ from Eq.(\ref{phi}) were used again. To check this calculations an expansion over $w=t/m^2$ in Eq.(\ref{fYF}) was engaged
\begin{equation}
\gamma\,=\,\int_0^1 \Big(\,\frac{\partial H(x,w)}{\partial w}\,\Big)_{w=0}dx.
\label{expw}
\end{equation}
Using (\ref{H}) it was verified that (\ref{expw}) reproduces (\ref{gamma}).

It is interesting that the imaginary part of the formfactor
$\mathrm{Im}\,f_\text{Y}(t)$ is able to change its sign. Fig. \ref{tri} illustrates this property. 
\begin{figure}[b]
\centering
\includegraphics[height=5.3 cm,keepaspectratio = true, 
%angle = -90
]{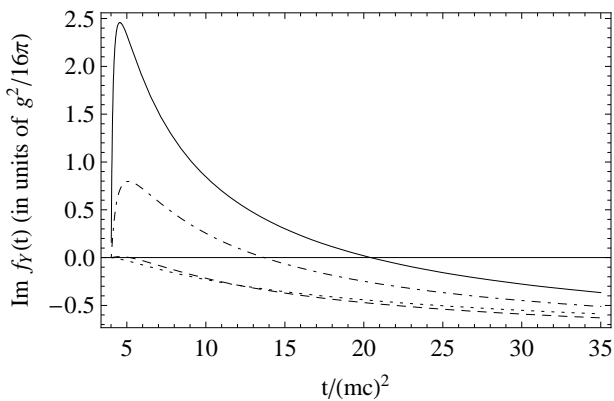}
\caption{
 \label{tri} The imaginary part of the formfactor $\mathrm {Im}\,f_\text{Y}(t)$ from Eq. (\ref{ImfY}) versus $t/m^2=k^2/m^2$ for different ratios of the fermion mass $m$
 and mass of the Higgs boson $m_\text{H}$. Solid, dot-dashed, dashed, and dotted lines correspond to  $m/m_\text{H}=5,~2.5,~1$, and $0.5$ respectively.}\end{figure}
\noindent
When $t$ is large then $\mathrm{Im}\,f_\text{Y}(t)$ is definitely negative since asymptotically $\mathrm{Im}\,f_\text{Y}(t) \rightarrow -g^2/(16\pi)$ when $t\rightarrow \infty$. However, Fig. \ref{tri} shows that if the ratio of masses $\mu=m_\text{H}/m$ is small, $\mu\le 2/\sqrt{3}\approx 1.15$ to be precise, then $\mathrm{Im}\,f_\text{Y}(t)$ at small $t$ exhibits a positive maximum, at some intermediate value of $t$ changes its sign, and only after that goes to its asymptotic value $-g^2/(16\pi)$. A magnitude of the peak at small $t$ depends on the mass ratio. The smaller $\mu=m_\text{H}/m$ the stronger the maximum.

To see origins of this interesting behaviour notice first of all that there is no restriction from the dispersion relation (\ref{u1}) on the sign of $\mathrm {Im}\,f_\text{Y}(t)$ since the initial $i$ and final $f$ states in Eq. (\ref{u1}) are different for the case at hand. Note further that the sign of the matrix element for the 
pair creation in Eq. (\ref{uu}) is not fixed, it depends on the direction of the momentum $\mathbf{p}$ of the fermion in the intermediate state and the spin states of the pair. As a result the sign of the function $\Phi(p)$, which plays a role of an integrand in Eq.(\ref{2ImfY}), depends on the orientation of the vector $\mathbf{p}$. It is foreseeable therefore that at different parameters the integral itself can be either positive, or negative. 

To make this point crystal clear one examines in detail the integrand $\Phi(p)$ in Eq.(\ref{2ImfY}), which according to Eq.(\ref{Phi(p)}) can be written as follows
\begin{equation}
\Phi(p)=\frac{g^2}{(2\pi)^4}\Big(1+
4m\frac{  \,\boldsymbol{q}\cdot \boldsymbol{\gamma} }
{\mathbf{q}^2+m_\text{H}^2}- 
\frac{4m^2+m_\text{H}^2}{\mathbf{q}^2+m_\text{H}^2}\,\Big).
\label{PhiCoul}
\end{equation}
It is taken into account here that in the center of mass coordinates of the fermion-antifermion pair the transferred momentum has only spacious components $q=(0,\mathbf{q})$. 
The first term in the brackets equals 1 and thus is positive, while the last term is always negative. 
At large $t$ the transferred momentum $\mathbf{q}$ is also large and therefore the first term dominates over the two others, which  makes the function $\Phi(p)$ positive and 
$\mathrm {Im}\,f_\text{Y}(t)$ in Eq.(\ref{ImfY}) negative, see Fig. \ref{tri} at large $t$.
At small $t$ the transferred momentum $\mathbf{q}$ is also small. If, in addition,
the Higgs mass is small as well, then the third term in (\ref{PhiCoul}) is dominant, the function $\Phi(p)$ under this condition is negative, while 
$\mathrm {Im}\,f_\text{Y}(t)$ in Eq.(\ref{ImfY}) is positive. One observes then the positive maximum in $\mathrm {Im}\,f_\text{Y}(t)$ at small $t$, see Fig. \ref{tri}.

Combining properties of the imaginary part of the formfactor illustrated by Fig. \ref{tri} with the dispersion relation (\ref{fYDR}), one deduces that the real part of the formfactor would necessarily show even more dramatic variations. However, we would not discuss  them in detail in general case. Instead, we restrict ourselves to a region of small $t$, which is important for applications. 

\begin{figure}[t]
\centering
\includegraphics[height=5.3 cm,keepaspectratio = true, 
%angle = -90
]{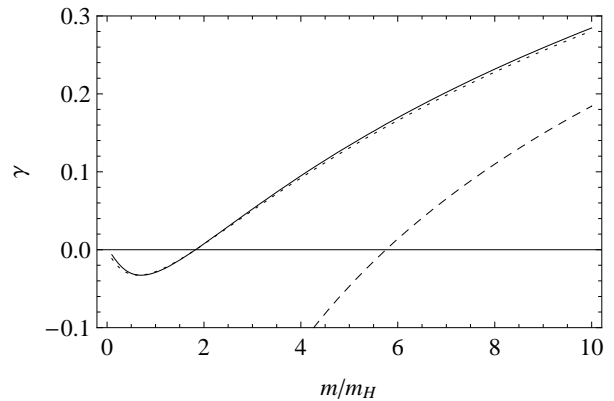}
\caption{
 \label{4etyre} The coefficient $\gamma$, which governs behaviour of the formfactor at small $t$ in Eq.(\ref{ft0}), versus  the ratio of 
the fermion and Higgs boson masses  $m/m_\text{H}=1/\mu$. Solid, dotted, and dashed
lines - precise expression (\ref{gamma}), fitting approximation (\ref{gfit}) and asymptote (\ref{gas})  respectively.}\end{figure}
\noindent

According to (\ref{ft0}) at small $t$ the formfactor is characterized by the only parameter $\gamma$, which depends though  on the mass ratio $m_\text{H}/m$ as specified in (\ref{gamma}). This dependence shown in Fig. \ref{4etyre} reveals an interesting feature, 
change of sign of $\gamma$ at $m=1.83\, m_\text{H}$.
In order to mimic this curious behaviour in simplest analytical terms it is convenient to introduce a fitting expression for $\gamma$ as follows
\begin{equation}
\gamma_\text{fit}\,=\,\frac{1}{3}\,\Big(\,\ln\frac{m+m_\text{H}}{m_\text{H}}-
\frac{7m}{4m+5m_\text{H}}\,\Big)~.
\label{gfit}
\end{equation}
It provides an accuracy of a few percent or better for all masses, see almost overlapping solid and dotted lines in Fig. \ref{4etyre}.
In contrast, an asymptotic relation 
\begin{equation}
\gamma_\text{asymp}\,=\,\frac{1}{3}\,\ln\frac{m}{m_\text{H}}-
\frac{7}{12}~,
\label{gas}
\end{equation}
which is derived from Eq.(\ref{gamma}) for $m \gg m_\text{H}$, converges slowly, see the dashed line in Fig. \ref{4etyre}.

Eq.(\ref{appear}) allows one to relate properties of $\gamma$ illustrated by Fig. \ref{4etyre} with behaviour of the imaginary part of the formfactor shown in Fig. \ref{tri}.  The rise of $\gamma$ at large $m/m_\text{H}$ is explained by the fact that the positive peak at small $t$ in $\mathrm {Im}\,f_\text{Y}(t)$ becomes stronger with increase of the mass ratio $m/m_\text{H}$; compare  the solid and	dot-dashed lines in Fig. \ref{tri}. This positive peak dominates in the integral in Eq.(\ref{appear}) and makes $\gamma$ positive and large. 
With the decrease of $m/m_\text{H}$ the peak in $\mathrm {Im}\,f_\text{Y}(t)$ is fading out, see the dashed and dotted lines in Fig. \ref{tri}. Correspondingly the negative continuum of the function $\mathrm {Im}\,f_\text{Y}(t)$ dominates the integral in Eq.(\ref{appear}) 
and makes $\gamma$ negative as well. 

We see that the sign and absolute value of $\gamma$ are closely related to the properties of $\mathrm {Im}\,f_\text{Y}(t)$, which shows different dependence on $t$ for different mass ratios $m/m_\text{H}$. As we know the latter fact is linked with processes in the $t$-channel related to creation and subsequent scattering of the fermion pair. Clearly, the mentioned properties of the formfactor arise from a conventional general relation, which binds the $s$ and $t$ channels, but it is rewarding  to observe its manifestations in this particular case.

We examined properties of the vertex correction presented by the diagram (b) in Fig. \ref{odin} for the case when the external fermion legs are on the mass shell. 
An important observation derived from this discussion is that at small momenta this correction does not depend on the fermion mass.
Later on, in Section 
\ref{Vertex correction to small momenta} 
we will discuss the same diagram in another interesting kinematic region, when all three external momenta are small.

\section{Fermion self-energy}
\label{Fermion-self-energy}
Consider the self-energy correction for fermions. From the diagram in Fig. \ref{odin} (c) we derive for it
\begin{equation}
\Sigma(p)\,=\,i\frac{g^2}{(2\pi)^4}\int\,G(q)D(q-p)\,d^4q~.
\label{Sigma}
\end{equation}
Using conventional renormalization conditions on the mass shell, when $\hat p=m$, 
\begin{equation}
\Sigma(p)\,=\,\frac{\partial \Sigma(p) }{ \partial p^{\mu} }\,=\,0~,
\label{S=S'=0}
\end{equation}
and fulfilling necessary integrations one finds
\begin{equation}
\Sigma(p)\,=\,\frac{g^2}{16\pi^2}\,\big(\,A(\rho)\,(\hat p-m)+B(\rho)\,m)\,\big)~.
\label{SAB}
\end{equation}
Here $\rho$ describes a deviation from the mass shell
\begin{equation}
\rho\,=\,(m^2-p^2)/m^2~,
\label{ro}
\end{equation}
and $A(\rho)$, $B(\rho)$ are defined as follows:
\begin{align}
&A(\rho)\,=\,a(\rho)-a(0)+2b^{\,\prime}(0)~,
\label{Aro}
\\
&B(\rho)\,=\,b(\rho)-b(0)~,
\label{Bro}
\\
&a(\rho)\,=\,\int_0^1\,x \,\ln Y(x,\rho)~dx~,
\label{arho}
\\
&b(\rho)\,=\,{\int_0^1}\, (1+x)\,\ln Y(x,\rho)~dx~,
\label{brho}
\\
&Y(x,\rho)\,=\,(1-x)^2+\mu^2 x+\rho\, x(1-x)~,
\label{yxro}
\end{align}
where as usual $\mu=m_\text{H}/m$. 
The integrals in Eqs.(\ref{arho}) and (\ref{brho}) can be expressed via elementary functions, but we stick to a more compact integral representation, which was introduced into the problem as an auxiliary tool in the Feynman parametrization of the denominators of the propagators, but proves be convenient to keep it in the final expressions.
To make this notation more comforting it is employed in
Appendix \ref{Mass operator in QED}  to present the known mass operator for spinor QED.

The renormalization conditions for $\Sigma(p)$ are implemented via the subtractions fulfilled in Eqs.(\ref{Aro}), (\ref{Bro}).
Simple algebraic calculations show that they allow the self-energy (\ref{SAB}) to be rewritten in the following form
\begin{equation}
\Sigma(p)\,=\,\frac{g^2}{16\pi^2}\,\frac{ (\hat p-m)^2 }{ m }\, \Upsilon~.
\label{SM}
\end{equation}
where 
\begin{align}
\Upsilon\,&=\,\frac{1}{\rho}\,\Big[ -B(\rho)+ 
\label{MassOper}
\\
&\frac{\hat p+m}{m}\,\Big(
2\frac{B(\rho)-B^{\prime}(0)\,\rho}{\rho}-A(\rho)+A(0)\Big)\,\Big]~.
\nonumber
\end{align}
On the mass shell, when $\hat p=m$ 
and $\rho=0$, $\Upsilon$  is finite, 
$
\Upsilon\,=\, -2A^{\prime}(0)-B^{\prime}(0)+2B^{\prime\prime}(0).
%\label{M0}
$
Consequently, Eq.(\ref{SM}) shows that in the vicinity of the mass shell $\Sigma(p)\propto (\hat p-m)^2$, which guarantees validity of the renormalization conditions (\ref{S=S'=0}).
\begin{figure}[t]
\centering
\includegraphics[height=5.2 cm,keepaspectratio = true, 
%angle = -90
]{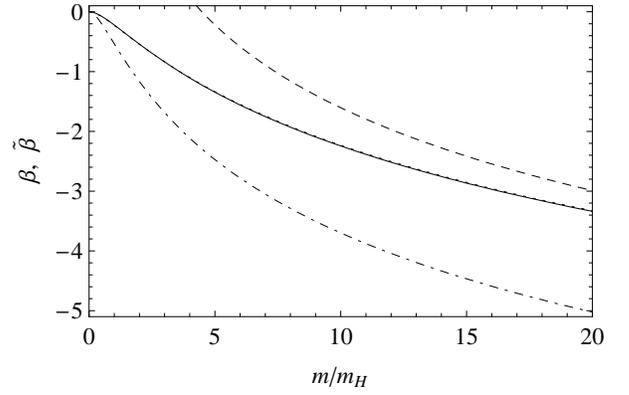}
\caption{
 \label{pyat'} The coefficients $\beta$ and $\tilde \beta$ in self-energy (\ref{S=bm}) and vertex (\ref{G00}) corrections versus the ratio of 
the fermion and Higgs boson masses  $m/m_\text{H}=1/\mu$. 
Solid, dotted and dashed lines - exact result (\ref{beta}), fitting approximation
(\ref{betaFit}), and asymptotic (\ref{betaAsym}) for $\beta$; dot-dashed line -  $\tilde \beta$ from Eqs.(\ref{tilde}),(\ref{beta}) }\end{figure}
\noindent

For applications related to the fermion bag it is interesting to consider the case of small momenta. Taking the limit $p=0$ we find from Eq.(\ref{SAB})
\begin{equation}
\Sigma(0)\,=\,\frac{g^2}{8\pi^2}\,\beta \,m~,
\label{S=bm}
\end{equation}
where $\beta=\big(B(1)-A(1)\big)/2$ is the function of the mass ratio $\mu=m_\text{H}/m$.
Straightforward calculations based on definitions of $A(\rho)$ and $B(\rho)$ in Eqs.(\ref{Aro}), (\ref{Bro}) allow one to find this function explicitly
\begin{align}
\beta\,=\,
&3-\mu^2+\Big(3-\frac{9}{2}\,\mu^2+\mu^4-\frac{1}{1-\mu^2}\Big)\ln \mu
\nonumber
\\
&-\frac{1}{2}\,(5-2\, \mu^2)\,(4-\mu^2)\,\phi(\,4-\mu^2)~.
\label{beta}
\end{align}
Here $\phi(w)$ is from Eq.(\ref{phi}).
Fig. \ref{pyat'} shows behaviour of $\beta$ as a function of $m/m_\text{H}=1/\mu$. 
It also shows the fitting approximation
\begin{equation}
\beta_\text{fit}\,=\,-2\,\ln \frac{m+3.32\,m_\text{H}}{m_\text{H}}+3 \,\frac{m+0.8\,m_\text{H}}{m+m_\text{H}}~,
\label{betaFit}
\end{equation}
which accuracy is few percent or better for all mass ratios, see almost overlapping solid and dotted lines in Fig. \ref{pyat'}, and the asymptote for $m\gg m_\text{H}$, which according to (\ref{beta}) reads 
\begin{equation}
\beta_\text{asymp}\,=\,-2\,\ln \frac{m}{m_\text{H}}+3~,
\label{betaAsym}
\end{equation}
and converges slowly, see the  dashed line in Fig. \ref{pyat'}. 

Eqs.(\ref{S=bm}) and (\ref{beta}) show that $\Sigma(0)$ is large and negative. Using (\ref{betaAsym}) one can write its asymptote at large $m$
\begin{equation}
\Sigma_\text{asymp}(0)\,=\, -\frac{1}{4\pi^2}\,\frac{m^3}{v^2} 
\Big(\ln \frac{m}{\,m_\text{H}}-\frac{3}{2}\,\Big)~.
\label{SmInfty}
\end{equation}
Eq. (\ref{SM}) allows one to estimate the value of the mass, which makes $\Sigma(0)$ compatible with the fermion mass $m$. Depending on the Higgs mass $m_\text{H} = 100-200$ Gev the equality $|\,\Sigma(0)\,|=m$ takes place at $m\approx 1300-1800$ Gev. For the fermion masses, which lie above this boundary the naive perturbation theory based on straightforward application of the one-loop approximation is not reliable. Hence the asymptotes in Eqs.(\ref{betaAsym}), (\ref{SmInfty}) represent merely the results of the one-loop approximation, which should deviate from true physical values.

\section{Vertex correction to fermion-Higgs interaction at small momenta}
\label{Vertex correction to small momenta}

Continue discussion of the vertex correction to the fermion-Higgs interaction represented by the diagram (b) in Fig. \ref{odin}. Previously, in Section \ref{mass-shell} this correction was calculated presuming that the fermion legs are on the mass shell, $p_1^2=p_2^2=m^2$. Let us investigate another important for applications situation when all external momenta of this diagram are small. This means that we presume that each component of each of the three vectors $k,p_1$ and $p_2$ are small compared to the fermion mass. 
\begin{equation}
|k_\mu|,\,|p_{1,\mu}|,\,|p_{2,\mu}|\ll m~,
\label{smallp}
\end{equation}
which implies that the fermion legs are well off the mass shell. 

Call $\Gamma(k,p)$ the contribution of the diagram (b) in Fig. \ref{odin}, where $k=p_1-p_2$, and $p=p_1+p_2$. According to Eq.(\ref{smallp}) we are interested in the limit $k,p\rightarrow 0$. The simplest way to evaluate $\Gamma(k,p)$ in this limit gives the Ward identity, which defines its behaviour at  $k=0$. For the considered fermion-Higgs interaction this identity reads
\begin{equation}
\Gamma(0,p)\, = \,g\,\frac{\partial}{\partial m}\,\Sigma(p)~.
\label{dsigma}
\end{equation}
Here $\Sigma(p)$ is the fermion self-energy, which in the one loop approximation 
is given by the diagram (c) in Fig. \ref{odin}. 
The validity of Eq.(\ref{dsigma}) is verified using an identity 
\begin{equation}
\frac{\partial}{\partial m}\,G(p)\,=\,G^2(p)~,
\label{G'=GG}
\end{equation}
which follows from (\ref{DG}).
In Eq.(\ref{dsigma}) it is presumed that in intermediate calculations the coupling constant $g$ and the fermion mass $m$ are treated as independent variables, and that the relation $m=gv$ from (\ref{g}) is implemented after the derivative in (\ref{dsigma}) is taken.

Note an important distinction of the considered kinematic region from the one in which the fermion legs are on the mass shell, see Section
\ref{mass-shell}. For the latter case the quantum correction to the vertex turns zero for small transferred momentum $\Gamma(0,p)=0$ when $p^2=m^2$. This
complies with the Ward identity (\ref{dsigma}) and the fact that in the vicinity of the mass shell $\Sigma(p)\propto  (\hat p-m)^2$, see Eq.(\ref{SM}). 
Condition $\Gamma(0,p)=0$ was used previously to normalize the formfactor in Eq.(\ref{fY=1}). This condition ensures that the lowest term of the expansion of the formfactor in powers of  $k^2$ is linear in $k^2$, see Eq.(\ref{ft0}).
In contrast, when the fermion legs are off the mass shell, as Eq.(\ref{smallp}) requires, the vertex remains nonzero at $k=0$.

Taking the limit of zero momentum $p=0$  in Eq.(\ref{dsigma}) and using Eq.(\ref{S=bm}) to represent  $\Sigma(0)$  we find for the vertex 
\begin{align}
&\Gamma(0,0)\, = \, \frac{g^3}{8\pi^2}\,\tilde \beta~,
\label{G00}
\\
&\tilde \beta \,=\,\frac{d}{dm}\big(\beta \,m\big)\,=\,\beta-\mu \,\frac{d\beta}{d\mu}~.
\label{tilde}
\end{align}
Fig. \ref{pyat'} shows $\tilde \beta$ found with the help of (\ref{beta}) as a function of $m/m_\text{H}=1/\mu$. From Eq. (\ref{betaAsym}) we derive its asymptotic behaviour at large fermion masses $m\gg m_\text{H}$ 
\begin{equation}
\tilde \beta_\text{asymp}\,=\,-2\,\ln \frac{m}{\,m_\text{H}}+1~,
\label{betaAsym1}
\end{equation}
Eqs.(\ref{G00}) and (\ref{tilde})  indicate that $\Gamma(0,0)$ is large and negative.
According to (\ref{betaAsym1}) its asymptote at large $m$ reads
\begin{equation}
\Gamma_\text{asymp}(0,0)\,=\, -\frac{1}{4\pi^2}\,\frac{m^3}{v^3} 
\Big(\ln \frac{m}{\,m_\text {H}}-\frac{1}{2}\,\Big)~.
\label{GmInfty}
\end{equation}
Combining this with (\ref{SmInfty}) and adopting the large-logarithm approximation one can write a simple relation 
\begin{equation}
\Sigma_\text{asymp}(0)\,\approx\,\Gamma_\text{asymp}(0,0)\,v~,
\label{GSv}
\end{equation}
which binds together the self-energy and vertex corrections for large fermion masses and small external momenta.
However, one has to remember the comment made at the end of Section \ref{Fermion-self-energy}.
For the fermion masses above the boundary $\sim 1300-1800$ Gev the perturbation theory is not reliable. Hence Eqs.(\ref{GmInfty}), (\ref{GSv}) refer to the results derived in the one-loop approximation, which may deviate from exact relations.

\section{Fermion loops for small momenta in Higgs legs}

It was mentioned in Section \ref{Formulation} that having in mind the bags with heavy fermions it is interesting to consider the situation, when the momenta in the external legs of a diagram are small. Consider now a fermion loop with several external Higgs legs presuming that the momenta running along these legs are small. Take for example the diagram with three Higgs legs shown in Fig. \ref{west'} (the legs represent scalars, 
therefore the odd number of legs is admissible).
\begin{figure}[htb]
\centering
\includegraphics[height=2.6 cm,keepaspectratio = true, 
%angle = -90
]{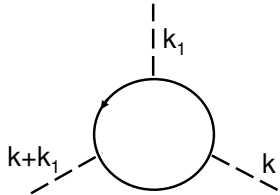}
\caption{
 \label{west'} An example of the fermion loop with several Higgs legs}\end{figure}
\noindent
Call its contribution $\Pi_{3}(k_1,k)$. Our purpose is to evaluate it at small momenta, when $\Pi_{3}(k_1,k)\approx \Pi_{3}(0,0)$. This goal can be achieved using the following Ward identity 
\begin{equation}
\Pi_{3}(0,k)\,=\,\frac{g}{2}\,\frac{\partial}{\partial m}\,P(t)~.
\label{G3=2P}
\end{equation}
Here $t=k^2$, and $P(t)$ is the polarization operator from Section 
\ref{Vacuum-polarization-fermions}.
Eq.(\ref{G3=2P}) is similar in nature to Eq.(\ref{G00}), and can be easily verified using the identity (\ref{G'=GG}). The coefficient $1/2$ in (\ref{G3=2P}) compensates for the fact that the derivative over the mass hits the product of the two fermion propagators, which are present in the integrand for $P(t)$.

Presuming that $t$ and $m_\text{H}$ are small compared with the fermion mass $m$ we can use for $P(t)$ its asymptotic expression Eq. (\ref{P2}). 
Remember that using the Ward identities one should  calculate the  derivative over the fermion mass presuming that $g$ and $m$ are two independent variables, and only after that relation $g=mv$ from (\ref{P2}) can be implemented. Therefore in Eq.(\ref{P2}) we need to take the middle expression, which treats $g$ and $m$ as separate variables. Substituting it into (\ref{G3=2P}) and setting there $k=0$ we find
\begin{equation}
\Pi_{3}(0,0)\,=\,
\frac{g^3}{80\pi^2}\frac{m_\text{H}^4}{m^3}\,=\,
\frac{1}{80\pi^2}\,\frac{m_\text{H}^4}{v^3}~.
\label{3leg}
\end{equation}
Observe that the fermion mass is canceled out in the final expression here
similarly to the way it was cancelated out in $P(t)$ in Eq.(\ref{P2}). 

The Ward identity Eq.(\ref{G3=2P}), which allowed us to derive $\Pi_{3}(0,0)$ from $P(0)$, can be generalized to cover the case of the fermion loop with an arbitrary number of the Higgs legs. Let us call $\Pi_{n}$  the contribution of the diagram in which the fermion loop has $n$ Higgs legs presuming that the momenta running along these legs are small compared to the fermion mass and therefore can be neglected. To clarify notation note that $\Pi_{2}= P(0)$, $\Pi_{3}=\Pi_{3}(0,0)$. Using Eq.(\ref{G3=2P}) one derives the following Ward identity
\begin{equation}
\Pi_{n+1}\,=\,\frac{g}{n}~\frac{\partial}{\partial m}\,\Pi_{n}~.
\label{Ward}
\end{equation}
Combining this with Eq.(\ref{3leg}) one finds
\begin{equation}
\Pi_{n}\,=\,
(-1)^{n+1}\frac{g^n}{80\pi^2}\frac{m_\text{H}^4}{m^n}\,=\,
(-1)^{n+1}\frac{1}{80\pi^2}\frac{m_\text{H}^4}{v^n}~.
\label{nleg}
\end{equation}
Setting here $n=2$ one recovers $P(0)$ from Eq. (\ref{P2}).

Observe that the fermion mass is canceled out from the final expression in Eq. (\ref{nleg}). Thus, we come to an interesting conclusion. A diagram with the fermion loop, which has any number of external Higgs legs, is not enhanced by the large coupling constant $g=m/v\gg 1$ provided momenta running along its legs are small.

%\vspace{0.5cm}
\section{Summary and discussion}
\label{conclusions}

An analysis of one-loop diagrams, which describe radiative corrections relevant to the fermion bags is presented. The bags in question are constructed from heavy fermions bound together by the Higgs boson exchange. Two areas, where the mass of a heavy fermion can belong are considered. Either the fermions are `moderately heavy', when $400\lesssim m\lesssim 1000$ Gev, or they are `very heavy', when $m>m$ Gev. 
In both kinematic regions one needs to study  small momenta ${\bf k}$, $|{\bf k}|\ll m$, which run along the external Higgs legs of the amplitudes. 
This is due to the size of the bag, which was found to be always larger than the Compton radius of the fermion \cite{Kuchiev:2008gt}. 

The difference between the two kinematic regions lies in the fact that for `moderately heavy' fermions one can use the mass shell approximation  $p^2=m^2$ for the external fermion legs of the diagram. In contrast, for `very heavy' fermions it is more appropriate to consider the limit of small momenta $p^\mu=0$ running along all fermion legs because the energy $\varepsilon=p_0$ of the fermion in the bag in this case is smaller than its mass, $\varepsilon\ll m$, while the spatial components ${\bf p}$ of the momentum are restricted by the mentioned large size of the fermion bag.

Remarkably, for `moderately  heavy' fermions it is found that none of the amplitudes, which define the radiative corrections, do depend on the fermion mass.
The large fermion mass is canceled out from the polarization operator and the Yukawa formfactor, see Eqs. (\ref{P2}) and (\ref{ft0}). Thus, these amplitudes are not enhanced by the large fermion mass in spite of the fact that the coupling constant, which is proportional to the mass, is not small, $g=m/v>1$.

This phenomenon can be understood without calculations using the following qualitative argument. Remember the non-relativistic Bethe-type approach to the Lamb shift problem in atoms. The electron self-interaction in this approach is estimated as $\propto \alpha \,\Delta U_e/m^2$, where $U_e$ is the electron potential energy. This estimate follows from the conventional QED electron formfactor. A similar structure appears in Eq. (\ref{ft0}). There we are dealing with a different object, the Yukawa formfactor, but it turns out that its properties are similar to the conventional electron formfactor in QED. The found Yukawa formfactor is proportional to  $\propto g^2 \,k^2/m^2$. The similarity with the Lamb shift problem becomes transparent if one draws parallels between  $g^2$ and $\alpha=e^2$ and identify  $k^2$ with $\Delta$. 

An important distinction between the considered case and the Lamb shift problem is that in our case the coupling constant $g=m/v$ is proportional to the fermion mass $m$. It is this fact that results in the mass cancellation in the final expression for the formfactor, which according to (\ref{ft0}) reads $\propto g^2 \,k^2/m^2=k^2/v^2$. 

The similar cancellation takes place for the  polarization operator $P(t)$ in (\ref{P2}). In that case this phenomenon can be supported by another simple qualitative argument. The renormalization conditions on the polarization operator $P(t)=P'(t)=0$ are formulated on the mass shell of the Higgs boson, for $t=m_\text{H}^2$, see  Eq.(\ref{ren}).  The Higgs mass plays the role of a small parameter, $m_\text{H}\ll m$. Consequently, $P(t)$ is suppressed for small momenta $t\sim m_\text{H}\sim v$. Precisely this suppression cancels out enhancement that comes from the large coupling constant $g=m/v$.

The mentioned cancellation of the large fermion mass indicates that in a bag of several `moderately heavy' fermions the radiative corrections are not enhanced by this mass. One should anticipate therefore that they are small. Probably they are very small having in mind the tiny numerical factor $1/(80 \pi^2)$ in Eq. (\ref{P2}) and a product of two small factors $1/(8\pi^2)$  and $\gamma$, which are present in (\ref{ft0}); remember that in the region considered $\gamma\sim 0.1 - 0.3$, see Fig. \ref{4etyre}.

The situation is more challenging for `very heavy' fermions, which is not surprising since in this region a strong coupling problem rises in all its might.
Correspondingly, the one-loop approximation, which was discussed in detail in the present work, is not sufficient. More sophisticated processes should give a substantial contribution. However, our discussion pinpoints an important feature of the problem, 
which is probably not restricted to the one-loop approximation and which leads to significant simplifications.

It was found that the fermion mass is canceled out from the final expressions for the diagrams with a fermion loop and several Higgs legs attached to it. 
This statement is valid provided the momenta running along the Higgs legs are small (which complies with the fact that the bag is large). 
The simplest diagram of this kind describes the polarization operator $P(t)$ mentioned above. 
The general case of the fermion loop with several Higgs legs can be treated by applying a chain of the Ward identities (\ref{Ward}) to the polarization operator.
Thus, the diagrams with the fermion loop are not enhanced by the fermion mass.  One has to expect therefore that their contribution to the radiative corrections is not significant.

However, for `very heavy' fermions the amplitudes, which are presented by the diagrams having external fermion legs, were found to be strongly enhanced by the large fermion mass.  
We described in detail two of them, the fermion self-energy correction $\Sigma$, and the vertex $\Gamma$ of the fermion-Higgs interaction. 
Both of them are enhanced by the fermion mass, $\propto m^3 \ln m$, if one restricts estimation to the one-loop approximation. 

It is tempting therefore to proclaim that the related radiative corrections are large, in agreement with the general pattern existing in literature, see Section \ref{intro}.
There are however indications, which do not comply with this assessment. In the region of `moderately heavy' fermions the radiative corrections are small and, importantly, independent of the fermion mass. It is difficult to reconcile this result with the expected 
large and mass-dependent corrections for `very heavy' fermions.
Another point to note is that there exists a sharp division between the two classes of diagrams.
The diagrams with closed fermion loops are found to be mass-independent and consequently are presumed small. In contrast, the diagrams with open fermion lines are strongly enhanced by the large fermion mass. This distinction looks artificial. One may anticipate that it appears only due to limitations of the one-loop approximation used.
It is worth reiterating here that when considering large masses, i.e. facing the strong coupling problem, it is dangerous to draw conclusions from the perturbation theory.
The reliable way would be to provide an accurate solution. We did not attempt to find it in the present work, but hopefully laid the groundwork for searching for it.

To summarize, properties of the amplitudes, which define radiative corrections in the fermion bag are discussed. The results are presented in a form, which makes them convenient for numerical calculations. For `moderately heavy' fermions these calculations are fulfilled in \cite{KF}, while for `very heavy' fermions the work ahead is more challenging.

\appendix

\section{One-loop polarization operator in QED}
\label{One-loop polarization operator in QED}

Assume that the fermions considered have the electric charge $e$. Then they give the following contribution to the imaginary part of ${\cal P}_\text{qed}(t)$, see e.g. section 115 of \cite{Berestetskii:1982},
\begin{equation}
\mathrm{Im}\,{\cal P}_\text{qed}(t)\,=\,-\frac{\alpha}{3}\,\left(
\frac{t-4m^2}{t} \right)^{1/2}\,(t+2m^2)~.
\label{PQED}
\end{equation}
Observe that at large $t\gg m^2$ there is a similarity between the two functions, $\mathrm{Im}\,{\cal P}_\text{qed}(t),\,\mathrm{Im}\,P(t)\propto t$, where $P(t)$ is the polarization operator from Section \ref{Vacuum-polarization-fermions}. In contrast, 
there is a substantial difference between them at the threshold $t\approx 4m^2$ . Here $\mathrm{Im}\,{\cal P}_\text{qed}(t)\propto (t-4m^2)^{1/2}$, while 
$\mathrm{Im}\,{P}(t)\propto (t-4m^2)^{3/2}$ is suppressed more strongly. An additional suppressing factor $(t-4m^2)$, which is present in $\mathrm{Im}\,{P}(t)$, arises from the matrix element $M_{in}=g\bar u(p)u(-p^{\,\prime})$, which turns out to be small at the threshold, $|M_{in}|\propto (t-4m^2)^{1/2}$. To verify this fact one takes explicit expressions for the Dirac spinors $u(p)$ and $u(-p^{\,\prime})$ in the cms and finds
\begin{equation}
\bar u(p)u(-p^{\,\prime})\,=\,-2 w^*\,(\boldsymbol{\sigma}\cdot \boldsymbol{p})\,
w^{\prime}~,
\label{uu}
\end{equation}
where $w$ and $w^{\prime}$ are two-component spinors, which describe spin states for the two fermions (not to be confused with the variable $w=t/m^2$ introduced previously). 
The matrix element in (\ref{uu}) complies with the parity conservation.  Under the inversion the momentum obviously changes sign, $\boldsymbol{p}\rightarrow -\boldsymbol{p}$, but in addition the product of the fermion and antifermion wave functions changes sign as well,
$w_\alpha^* w_\beta^{\prime}\rightarrow 
-w_\alpha^*w_\beta^{\prime}$, which ensures that $\bar u(p)u(-p^{\,\prime})$ is even, as it should be.
Eq.(\ref{uu}) explicitly shows that the amplitude of the pair creation is suppressed at the threshold $|M_{in}|\propto |{\bf p}|\propto (t-4m^2)^{1/2}$. This fact, in turn, justifies an additional factor $t-4m^2$, which distinguishes  $\mathrm{Im}\,P(t)$ from $\mathrm{Im}\,{\cal P}_\text{qed}(t)$ at the threshold.

To complete comparison between $P(t)$ and ${\cal P}_\text{qed}(t)$ let us write down an analytical expression for ${\cal P}_\text{qed}(t)$ in terms similar to those used for $P(t)$ in (\ref{Pf}). From the imaginary part (\ref{PQED}) one recovers the polarization operator
\begin{equation}
{\cal P}_\text{qed}(t)\,=\,\frac{2\alpha}{3\pi}\,m^2\,\Big((w+2)\,\phi(w)-2-\frac{5}{6}\,w\,\Big)~.
\label{PFQED}
\end{equation}
The second and third terms in the brackets here implement the renormalization conditions.
%Verifying them one takes into account that 
%$F_\text{qed}(0)=2$ and $F_\text{qed}^{\,\prime}(0)=5/6$. 
One verifies that ${\cal P}_\text{qed}(t)$ from Eq.(\ref{PFQED}) coincides with the well-known expression for the one-loop polarization operator in QED, see e.g. section 113 of \cite{Berestetskii:1982}.

\section{Vacuum polarization by Higgs field}
\label{Vacuum polarization by Higgs field}
\begin{figure}[t]
\centering
\includegraphics[height=1.6 cm,keepaspectratio = true, 
%angle = -90
]{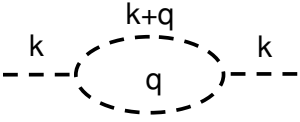}
\caption{
 \label{dva} Vacuum polarization by the Higgs field.}\end{figure}
\noindent
Consider the vacuum polarization due to the Higgs field shown in Fig. \ref{dva}.
It has some similarity with the polarization produced by fermions, see Fig. \ref{odin} (a). The difference is that 
the vertexes of the fermion loop in the diagram in Fig. \ref{odin} (a) are enhanced by the large fermion mass, while the the Higgs loop in 
Fig. \ref{dva} acquires no enhancement from the fermion mass. However, we saw in Eq.(\ref{P2}) that the fermion mass is canceled out from the final expression for the polarization operator which originates from Fig. \ref{odin} (a).  It makes sense therefore to include in our discussion the vacuum polarization produced by the diagram Fig. \ref{dva}.

Let us call $P_\text{H} (t)$ the contribution of this diagram to the polarization operator which describes propagation of the scalar field. The Feynman rules specify that it reads
\begin{equation}
P_\text{H} (t)\,=\,i\,\frac{ \nu\,\lambda_3^2}{(2\pi)^4}\,\int\,D(q)D(k+q)\,d^4q~.
\label{SDD}
\end{equation}
Here the usual factor $\nu=1/2$ in front takes into account the Bose statistics as well as the fact that the nonlinear $\propto \varphi^3$ coupling was defined in Eq.(\ref{LNL}) with the factor $\lambda_3/3!\,$. After the renormalization of Eq.(\ref{SDD})  straightforward calculations give
\begin{equation}
P_\text{H} (t)\,=\,\frac{\nu\,\lambda_3^2}{8\pi^2}~[\,\phi(z)-\phi(1)-(z-1)\,\phi'(1)\,]~.
\label{S(t)}
\end{equation}
Here $z=t/m_\text{H}^2=k^2/m_\text{H}^2$ and the function $\phi(z)$ was defined in Eq.(\ref{phi}).
Eq.(\ref{S(t)}) ensures that the renormalization conditions
$S(m_\text{H}^2)\,=\,S^{\,\prime}(m_\text{H}^2)\,=\,0$
are satisfied. 

For $t>4m_\text{H}^2$ there exists an imaginary part of $P_\text{H} (t)$. Eq.(\ref{S(t)}) shows that it  equals
\begin{equation}
\mathrm{Im}\,P_\text{H} (t)\,=\,-\frac{\nu\,\lambda_3^2}{16\pi}\,\Big(\,\frac{t-4m_\text{H}^2}{t}\,\Big)^{1/2}~.
\label{ImS}
\end{equation}
Let us verify that it complies with the unitarity condition (\ref{u1}). Take in the latter one $i=f$ as a state with one virtual Higgs boson, so that 
$M_{ii}=-P_\text{H} (t)$. Consider $n$ as a state with two real bosons. Then $M_{in}=-\sqrt{2}\cdot3\cdot\lambda_3^2/3!\,$. Here the factor $-\lambda_3^2/3!$ comes from the nonlinear $\varphi^3$ term in the Lagrangian (\ref{LNL}), the factor 3 takes into account that the virtual boson can be attributed to any one of the three operators present in the vertex $\varphi^3$, while the factor $\sqrt{2}$ accounts for the Bose statistics which governs the two bosons in the intermediate state $n$. There is no need for summation over spins in (\ref{u1}), the integration over the angles of the intermediate boson is trivial, gives $4\pi$, and one immediately verifies that the unitarity condition results in the same imaginary part of $P_\text{H} (t)$ as the one in (\ref{ImS}). Thus, the expression (\ref{S(t)}) for $P_\text{H} (t)$  complies with the unitarity condition.

Consider the limiting cases. For large momenta, when $|t|\gg m_\text{H}^2$, Eq.(\ref{S(t)}) gives 
\begin{equation}
P_\text{H} (t)\,\approx\, 
\frac{\nu \lambda_3^2}{16\pi^2} \,\ln (-t/m_\text{H}^2-i0\,)~.
\label{Slarget}
\end{equation}
%\begin{equation}
%P_\text{H} (t)\rightarrow 
%\frac{\nu \lambda_3^2}{16\pi^2}\times\left\{
%\begin{array}{ll}
%(\,\ln (\,t/m_\text{H}^2\,)-i\pi\,),& \quad t>0 
%\\
%~\ln (\,|\,t\,|/m_\text{H}^2\,),        &\quad t<0 
%\end{array}
%\right.
%\label{Slarget}
%\end{equation}
For small $|t|\ll m_\text{H}^2$ one derives from Eq.(\ref{S(t)}) that
\begin{equation}
P_\text{H} (t)\,\approx\,\frac{\nu \lambda_3^2}{8\pi^2}\,(s_0+s_1 z )~,
%+s_2 z^2+s_3 z^3+\dots
\label{Ssmallt}
\end{equation}
where $z=t/m_\text{H}^2$ and the coefficients are
$s_0=3/2-5\pi/(6\sqrt{3}) \approx -0.011$,
$s_1=-7/12+\pi/(3\sqrt{3})\approx 0.021$.

Compare the contribution, which comes from the Higgs field and from fermions into the vacuum polarization. Having in mind applications to the fermion bags we can assume that 
$t\sim 1/R^2$, where $R$ is the radius of the bag. Using the estimate  $R\sim R_0= (N/2\pi)^{1/2}/v$ from  \cite{Kuchiev:2008gt} where $N$ is the number of fermions in the bag, $N\le 12$, we estimate $t$ as follows, $t\sim t_0= v^2 N/(2\pi)$. Taking into account that $t_0/m^2\ll 1$ one can approximate the fermion contribution to the vacuum polarization $P(t_0)$ via its asymptotic form in Eq.(\ref{P2}). The Higgs contribution to the vacuum polarization should be taken from the exact expression (\ref{SDD}) because $t_0/m_\text{H}^2$ is not necessarily small. We evaluate then numerically $P_\text{H}(t_0)$ and $P(t_0)$ considering them as functions of $m_\text{H}$ for $100\le m_\text{H}\le 200$ Gev. Using this procedure we find that in the given interval of the Higgs masses the vacuum polarization by fermions  $P(t_0)$ exceeds the Higgs polarization $P_\text{H}(t_0)$ by a factor of at least 2 or more. We conclude that in the fermion bags the vacuum polarization produced by fermions is more important than the polarization due to the Higgs field, though this distinction is only numerical.

\section{Mass operator in QED}
\label{Mass operator in QED}

Consider the well known one-loop approximation for the mass operator in spinor electrodynamics. In the Feynman gauge for the photon propagator it can be presented as follows
\begin{equation}
\Sigma_\text{qed}(p)\,=\,i\,\frac{-4\pi e^2}{(2\pi)^4}\int \gamma^\mu G(q)\gamma_\mu D_\text{ph}(q-p)\,d^4q~.
\label{massOp}
\end{equation}
Here $D_\text{ph}(q)=1/(q^2-\lambda^2)$ is the photon propagator, and $\lambda$ is the infrared cut-off parameter. Compare this expression with the self-energy from Eq.(\ref{Sigma}). Observe that (\ref{massOp}) can be derived from (\ref{Sigma}) if we make in the latter the following substitutions: $g^2\rightarrow 4\pi e^2$, $m_\text{H}\rightarrow \lambda$, and 
$G(q)\rightarrow -\gamma^\mu G(q)\gamma_\mu=2(\hat q-2m)/(q^2-m^2) $. Straightforward calculations show that this similarity allows one to present the mass operator (\ref{massOp}) in the form similar to Eqs.(\ref{SAB})-(\ref{yxro}),
\begin{align}
\Sigma_\text{qed}(p)=\frac{\alpha}{2\pi}\big(\,{\cal A}_\text{qed}(\rho)\,(\hat p-m)+
{\cal B}_\text{qed}(\rho)\,m\,\big).
\label{MQED}
%\\
%&A_\text{qed}(\rho)\,=\,a(\rho)-a(0)+2b^\prime_\text{qed}(0)~,
%\\
%&B_\text{qed}(\rho)\,=\,b_\text{qed}(\rho)-b_\text{qed}(0)~,
%\\
%&b_\text{qed}(\rho)\,=\,3a(\rho)-2b(\rho)~.
\end{align}
The functions ${\cal A}_\text{qed}(\rho)$ and ${\cal B}_\text{qed}(\rho)$ here are derived from $A(\rho)$ and $B(\rho)$ correspondingly using the substitution $b(\rho)\rightarrow b_\text{qed}(\rho)=3a(\rho)-2b(\rho)$ in the right-hand sides of Eqs. (\ref{Aro}) and (\ref{Bro}), which define these functions.  One verifies that Eq.(\ref{MQED}) reproduces the mass operator found by Karplus and Kroll in \cite{Karplus-Kroll}, see also \cite{Berestetskii:1982}.

%\vspace{2cm}

This work is supported by the Australian Research Council

%\vspace{1 cm}


\begin{thebibliography}{10}



\bibitem{Vinciarelli:1972zp}
P.~Vinciarelli,
\newblock Lett. Nuovo Cim. {\bf 4S2}, 905 (1972).

\bibitem{PhysRevD.9.2291}
T.~D. Lee and G.~C. Wick,
\newblock Phys. Rev. D {\bf 9}, 2291 (1974).

\bibitem{Chodos:1974je}
A.~Chodos, R.~L. Jaffe, K.~Johnson, C.~B. Thorn, and V.~F. Weisskopf,
\newblock Phys. Rev. {\bf D9}, 3471 (1974).

\bibitem{Creutz:1974bw}
M.~Creutz,
\newblock Phys. Rev. {\bf D10}, 1749 (1974).

\bibitem{Bardeen:1974wr}
W.~A. Bardeen, M.~S. Chanowitz, S.~D. Drell, M.~Weinstein, and T.-M. Yan,
\newblock Phys. Rev. {\bf D11}, 1094 (1975).

\bibitem{Giles:1975gy}
R.~Giles and S.~H.~H. Tye,
\newblock Phys. Rev. {\bf D13}, 1690 (1976).

\bibitem{Huang:1975ih}
K.~Huang and D.~R. Stump,
\newblock Phys. Rev. {\bf D14}, 223 (1976).

\bibitem{PhysRevD.15.1694}
R.~Friedberg and T.~D. Lee,
\newblock Phys. Rev. D {\bf 15}, 1694 (1977).

\bibitem{PhysRevD.25.1951}
R.~Goldflam and L.~Wilets,
\newblock Phys. Rev. D {\bf 25}, 1951 (1982).

\bibitem{PhysRevLett.53.2203}
R.~MacKenzie, F.~Wilczek, and A.~Zee,
\newblock Phys. Rev. Lett. {\bf 53}, 2203 (1984).

\bibitem{PhysRevD.32.1816}
L.~R. Dodd and M.~A. Lohe,
\newblock Phys. Rev. D {\bf 32}, 1816 (1985).

\bibitem{Khlebnikov:1986ky}
S.~Y. Khlebnikov and M.~E. Shaposhnikov,
\newblock Phys. Lett. {\bf B180}, 93 (1986).

\bibitem{Anderson:1990kb}
G.~W. Anderson, L.~J. Hall, and S.~D.~H. Hsu,
\newblock Phys. Lett. {\bf B249}, 505 (1990).

\bibitem{MacKenzie:1991xg}
R.~MacKenzie,
\newblock Mod. Phys. Lett. {\bf A7}, 293 (1992).

\bibitem{Macpherson:1993rf}
A.~L. Macpherson and B.~A. Campbell,
\newblock Phys. Lett. {\bf B306}, 379 (1993), hep-ph/9302278.

\bibitem{Johnson:1986xz}
R.~Johnson and J.~Schechter,
\newblock Phys. Rev. {\bf D36}, 1484 (1987).

\bibitem{Dimopoulos:1990at}
S.~Dimopoulos, B.~W. Lynn, S.~B. Selipsky, and N.~Tetradis,
\newblock Phys. Lett. {\bf B253}, 237 (1991).

\bibitem{Bagger:1991pg}
J.~A. Bagger and S.~G. Naculich,
\newblock Phys. Rev. Lett. {\bf 67}, 2252 (1991).

\bibitem{Farhi:1998vx}
E.~Farhi, N.~Graham, P.~Haagensen, and R.~L. Jaffe,
\newblock Phys. Lett. {\bf B427}, 334 (1998), hep-th/9802015.

\bibitem{Farhi:2003iu}
E. ~Farhi, N. ~Graham,  R. L. ~Jaffe, V. ~Khemani, and H. ~Weigel, ,
Nucl. Phys. {\bf B665}, 623 (2003); hep-th/0303159.

\bibitem{Froggatt:etal}
C.~D. Froggatt and H.~B. Nielsen,
\newblock Phys. Rev. {\bf D80}, 034033 (2009), 0811.2089;
 D. Froggatt, H. B. Nielsen, and L. V. Laperashvili, Int. J. Mod. Phys. A20, 1268 (2005), hep-ph/0406110; C. D. Froggatt and H. B. Nielsen, Surveys High Energ. Phys. 18, 55 (2003), hep-ph/0308144.

%\bibitem{Froggatt:2008hc}
%C.~D. Froggatt and H.~B. Nielsen,
%\newblock Phys. Rev. {\bf D80}, 034033 (2009), 0811.2089.

\bibitem{Kuchiev:2008fd}
M.~Y. Kuchiev, V.~V. Flambaum, and E.~Shuryak,
\newblock Phys. Rev. {\bf D78}, 077502 (2008), 0808.3632.

\bibitem{Richard:2008uq}
J.-M. Richard,
\newblock Few Body Syst. {\bf 45}, 65 (2009), 0811.2711.

\bibitem{Kuchiev:2010ux}
M.~Y. Kuchiev,
\newblock Phys. Rev. D (accepted 2010), 1009.2012.

\bibitem{Crichigno:2009kk}
M.~P. Crichigno and E.~Shuryak,
\newblock (2009), 0909.5629.

\bibitem{Crichigno:2010ky}
M.~P. Crichigno, V.~V. Flambaum, M.~Y. Kuchiev, and E.~Shuryak,
\newblock (2010), 1006.0645.

\bibitem{Kuchiev:2008gt}
M.~Y. Kuchiev, V.~V. Flambaum, and E.~Shuryak,
\newblock Phys. Lett B.  (2010), 0811.1387.


\bibitem{Bagger:1985ig}
J.~Bagger, S.~Dimopoulos, and E.~Masso,
\newblock Phys. Rev. Lett. {\bf 55}, 920 (1985).

\bibitem{Babu:1985ut}
K.~S. Babu and E.~Ma,
\newblock Phys. Rev. Lett. {\bf 55}, 3005 (1985).

\bibitem{Berestetskii:1982}
V.~B. Berestetskii, L.~P. Pitaevskii, and E.~M. Lifshitz,
Quantum Electrodynamics, Course of theoretical physics, Vol. 4, 2nd ed. 
\newblock (2002),
\newblock %Oxford, England: 
Pergamon. %Press.

\bibitem{Karplus-Kroll}
R.~ Karplus and N.~M.~ Kroll,
Phys. Rev. {\bf 77}, 536 (1950).

\bibitem{KF}
M.~Yu.~Kuchiev and V.~V.~Flambaum, submitted to arXiv [hep-ph] 04.12.2010.

\end{thebibliography}
\end{document}